\documentclass[10pt]{iopart}
\pdfoutput=1
\usepackage{microtype}
\usepackage[utf8]{inputenc}
\usepackage{amsmath}
\usepackage{amssymb}
\usepackage{graphicx}
\usepackage{multirow}
\usepackage{color}
\usepackage{cite}

\graphicspath{{figs/}}

\newcommand{\Lf}{\mathcal{L}}

\newcommand{\p}[1]{ \left( #1 \right) }

\newcommand{\mE}{ \mathcal E }
\newcommand{\mG}{ \mathcal G }
\newcommand{\mR}{ \mathcal R }

\newcommand{\mA}{ \mathcal A }

\newcommand{\Prob}{P}
\newcommand{\mId}{ \mathbb I }
\newcommand{\Tra}{ \mathrm {tr}\, }

\newcommand{\mP} {\mathcal{P}}

\newcommand{\Esp}[1]{\mathbb E\left[#1\right]}

\newcommand{\iid}{\mathrm{i.i.d} }

\newcommand{\R}{\mathbb R}

\newcommand{\Res}{\mathrm{Res}}
\newcommand{\Col}[1]{w_{#1}^*}

\begin{document}
\title[Large deviations for random variables described by a matrix product ansatz]{Large deviations for correlated random variables described by a matrix product ansatz}
\author{Florian Angeletti,$^{1,2}$ Hugo Touchette,$^{1,2}$ Eric Bertin,$^{3,4}$ and Patrice Abry$^{3}$}

\address{
$^1$\ National Institute for Theoretical Physics (NITheP), Stellenbosch 7600, South Africa\\
$^2$\ Institute of Theoretical Physics, University of Stellenbosch, Stellenbosch 7600, South Africa\\
$^3$\ Laboratoire de physique, UMR CNRS 5672, \'Ecole Normale Sup\'erieure de Lyon, 46 All\'ee d'Italie, Lyon 69007, France \\
$^4$ Laboratoire Interdisciplinaire de Physique, Universit\'e Joseph Fourier Grenoble, CNRS UMR 5588, BP 87, F-38402 Saint-Martin d'H\`eres, France\\
}

\eads{
\mailto{florianangeletti@sun.ac.za},
\mailto{htouchette@sun.ac.za},
\mailto{eric.bertin@ujf-grenoble.fr},
\mailto{patrice.abry@ens-lyon.fr}
}

%\maketitle
%\tableofcontents

\begin{abstract}
We study the large deviations of sums of correlated random variables described by a matrix product ansatz, which generalizes the product structure of independent random variables to matrices whose non-commutativity is the source of correlations. We show with specific examples that different large deviation behaviors can be found with this ansatz. In particular, it is possible to construct sums of correlated random variables that violate the Law of Large Numbers, the Central Limit Theorem, as well as sums that have nonconvex rate functions or rate functions with linear parts or plateaux.
\end{abstract}

\date{\today}

\section{Introduction}
 
The study of stationary states and fluctuations of nonequilibrium systems using concepts and methods from large deviation theory has become an active topic in statistical mechanics \cite{bertini2007,derrida2007,touchette2009,harris2013}, following the successful application of this theory to equilibrium systems \cite{ellis1985,oono1989,ellis1995,ellis1999}. For both types of systems, it is known that the calculation of rate functions, the central object of large deviation theory characterizing the likelihood of fluctuations, is in general equivalent to, and therefore as difficult as, calculating partition functions \cite{touchette2009}. From this point of view, the simplest systems for which large deviations can be obtained are systems of independent particles, which correspond in probabilistic terms to independent and identically distributed ($\iid$) random variables. Next come linear chains of particles interacting via first-neighbor potentials, which can be mapped to Markov chains. In this case, results such as the G\"artner-Ellis Theorem can be used to obtain rate functions by solving an eigenvalue problem, which is essentially a transfer operator problem \cite{ellis1985,touchette2009,harris2013}.

Large deviations have been obtained for other models of correlated systems: e.g., random and Gibbs fields \cite{georgii1988,follmer1988b,georgii1993,eizenberg1994}, random matrices \cite{anderson2010,vivo2007,guionnet2004,dean2006,katzav2010}, random walks in random environments \cite{greven1994,hollander2000,zeitouni2006},  hidden Markov processes \cite{ephraim2002,kargin2007,fabiola2012}, and processes that explicitly evolve in non-Markovian way \cite{harris2009}. However, as one goes beyond Markov processes the applications of large deviation techniques becomes very difficult and often leads to non-explicit results. Not much, in particular, is known on the application of the G\"artner-Ellis Theorem for sums of strongly correlated random variables -- a problem mirrored again in statistical mechanics in the difficulty of calculating partition functions for strongly correlated particles.

The goal of this contribution is to present a class of correlated random variables, defined via a matrix product ansatz, for which the large deviations of the sample mean can be obtained explicitly. This class of random variables was proposed recently in \cite{Angeletti2012:ICASSP,Angeletti2013:PiMat} as a generalization of recent results on stationary distributions of interacting particle models, in particular, the asymmetric exclusion process, which has been actively studied recently \cite{Hakim83,Crampe2011,Lazarescu2011,Lazarescu2013,DerridaASEP93,Mallick97,Evans07}. The basis of this ansatz, presented in the next section, is to express the joint probability distribution of a sequence of random variables as a product of matrices whose non-commutativity is the source of correlations between these random variables.

Our goal here is to obtain large deviation results for the sample mean of these ``matrix-correlated'' random variables using the G\"artner-Ellis Theorem, and to show that the application of this theorem leads in this context to a simple product structure for generating functions, similar to that of $\iid$ random variables, but involving matrices rather than scalar functions. The difference is crucial as it leads to large deviation behaviors that do not arise for $\iid$ random variables and ergodic finite Markov chains. We will show, for example, that sample means having nonconvex rate functions are possible for matrix-correlated random variables, as are rate functions with linear parts or plateaux. These examples are related, as will be explained, to extensions of the Law of Large Numbers and Central Limit Theorem, recently studied in \cite{Angeletti2013:MatSum:Letter}. Here we validate and complement this study from the point of view of large deviation theory.

\section{Definitions and model }
\label{sec:RVMR}

We study the sum
\begin{equation}
 S_n = \sum_{i=1}^{n} X_i
\end{equation}
of $n$ random variables $X_1,\ldots,X_n$ whose correlations are described by the joint probability density function (pdf) $\Prob(x_1, \dots, x_n)$.\footnote{We consider throughout real random variables, but discrete random variables are also possible.} The general model of correlation that we consider is defined by the following form for the joint pdf, referred to as the \emph{matrix product ansatz}: 
\begin{equation} \label{eq:def}
\Prob(x_1, \dots, x_n) = \frac{1}{\Lf\p{\mE^n}} \Lf\p{\mR\p{x_1} \dots \mR\p{x_n} }
\end{equation}
where $\mR(x)$ is a positive $d \times d$ matrix function, $\Lf$ is a linear form defined by
\begin{equation} 
\Lf\p{M} = \Tra {\mA^T M }
\end{equation} 
with $\mA$ a positive $d \times d$ matrix, and $\mE$ is the \emph{structure} matrix defined by $\mE = \int \mR(x) dx$ and such that $\Lf\p{\mE^n} \ne 0$.

This ansatz was proposed and studied in \cite{Angeletti2013:PiMat,Angeletti2013:MatSum:Letter}, following similar forms of joint pdfs appearing in the context of nonequilibrium particle models, such as the exclusion process \cite{DerridaASEP93,Mallick97,Evans07}.
Similar matrix product state ansatz also appear in quantum many-body physics \cite{Verstraete2008, Lesanovsky2013}.
 The main property of this model is that the correlations in $X_1,\dots,X_n$ are controlled by the structure matrix $\mE$. To be more precise, let $\nu_1,\dots,\nu_m$ with $m\leq d$ denote the \textit{distinct} eigenvalues of $\mE$, ordered in descending order of their real part, and let $J_{k,l}$ be the $l$-th Jordan block associated with the eigenvalue $\nu_k$  in a Jordan basis $B$ of $\mE$:
\begin{equation}
 \mE= B \begin{pmatrix} J_{1,1} & &0\\ & \ddots & \\0& & J_{m, r} \end{pmatrix} B^{-1},
\quad J_{k,l} =
\begin{pmatrix}
\nu_k& 1    	&    		&0 \\
   	& \ddots 	& \ddots 	&  \\
   	&     		& \ddots 	& 1 \\
0  	&    		&    		& \nu_k \\
\end{pmatrix}.
\end{equation}
From the Perron-Frobenius Theorem for non-negative matrices,  the dominant eigenvalue $\nu_1$ of $\mE$ is positive real. The Jordan blocks $J_{k,l}$ determine the type of correlation in $X_1,\ldots, X_n$ according to the following cases (which are not mutually exclusive)\cite{Angeletti2013:PiMat}:

\begin{enumerate}
\item If there is at least one block $J_{k,l}$ with $k\ne1$ (i.e., there exists at least one non-dominant eigenvalue), then the sequence of random variables $X_1,\ldots,X_n$ exhibits exponential, \emph{short-range} correlation, as is typically the case for ergodic Markov chains. An example of matrix $\mE$ that falls into this case are irreducible aperiodic matrices, such that, for some $k>0$, $\p{\mE^k}_{i,j} > 0$ for any $i,j$.

\item If the dominant eigenvalue $\nu_1$ has more than one Jordan blocks, then the sequence $X_1,\ldots,X_n$ exhibits (generically non-zero) constant correlations, as arises, for example, in non-ergodic Markov chains.
The identity matrix $\mId$ provides the simplest subclass of matrices with this kind of Jordan normal form.

\item If there is at least one block $J_{1,l}$ associated with $\nu_1$ with dimension greater than 1, then $X_1,\ldots,X_n$ exhibits
polynomial long-range correlation in the sense that $\Esp{X_k X_l} \approx Q(k/n, l/n)$ with $Q(X,Y)$ a polynomial function. 
The simplest example of such matrices are the so-called linear irreversible subclass
\begin{equation} \label{eq:RVMR:mEIU}
\mE = \mId + U,\quad U=\begin{pmatrix}0 & 1 &  0\\ \vdots & \ddots& 1 \\ 0 & \cdots & 0 \end{pmatrix} .
\end{equation}
\end{enumerate}

As shown in \cite{Angeletti2013:PiMat}, these three cases can also be understood by noting that the product ansatz admits a representation in terms of hidden Markov chains, consisting of a hidden Markov chain layer $\Gamma_1,\ldots,\Gamma_{n+1} \in \{1,\dots ,d \}$ determining the visible layer $X_1,\ldots, X_n$. The transition matrix of the $\Gamma$ Markov chain is obtained from the structure matrix $\mE$ according to
\begin{align}
\label{eqn:trans0}
P(\Gamma_1 = i,\Gamma_{n+1} = f) &= \mA_{if} \frac{(\mE^n)_{if}}{\Lf(\mE^n)} \; , \\
 \label{eq:transition}
P(\Gamma_{k+1}= j | \Gamma_{k}=i, \, \Gamma_{n+1} = f ) &= \mE_{ij} \frac{(\mE^{n-k})_{jf}}{(\mE^{n-k+1})_{if}} \; .
\end{align}
This Markov chain is non-homogeneous and nonstandard, due to the dependence on the final state $\Gamma_{n+1}$.
In particular for $k=n$, the transition rate  $P(\Gamma_{k+1}= j | \Gamma_{k}=i, \, \Gamma_{n+1} = f )$
equals $1$ if $j=f$ and $0$ otherwise, meaning that the last step of $\Gamma$ is deterministic.
For a given sequence $\Gamma_1,\ldots,\Gamma_{n+1}$, the random variables $X_1,\ldots,X_n$ are then independent but non-identically distributed, 
with a pdf depending on $\Gamma$:
\begin{equation}\label{eq:X|gamma}
P(x_1,\ldots,x_n|\Gamma) = \prod_{k=1}^n \mP_{\Gamma_k \Gamma_{k+1}}(x_k),
\end{equation}
where
\begin{equation}
\mP_{i,j}(x)=\frac{\mR_{i,j}(x)}{\mE_{i,j}}
\label{Pdef1}
\end{equation}
represents the conditional pdf associated with transitions from states $i$ to $j$ in the hidden layer. With this result, the exponentially-decaying and constant correlation cases, mentioned above, can be understood as arising from similar correlations at the hidden Markov chain $\Gamma$ level, while the polynomial correlation case is more unusual and originates from the non-homogeneous nature of $\Gamma$. At this point, it is important to note that the states of the Markov chain $\Gamma$ are not the values of the random variables $X_1,\ldots,X_n$: the former has $d$ discrete states, as seen from (\ref{eqn:trans0}) and (\ref{eq:transition}), while the $X_i$'s are again real random variables with joint pdf (\ref{eq:X|gamma}).

The hidden Markov chain representation of matrix-correlated random variables is useful for designing and synthesizing random variables with prescribed statistical properties, such as fixed marginal pdfs, correlation functions, or higher order dependencies \cite{Angeletti2013:PiMat}. It can also be used to derive analogues of the Law of Large Numbers and the Central Limit Theorem \cite{Angeletti2013:MatSum:Letter}. In this case, it has been found that non-standard limit laws appear for random variables with polynomial or constant correlation, illustrating the difference between these two kinds of long-range correlation and the exponentially-decaying kind, for which standard laws typically apply.

Here, we are interested in extending these results by studying the large fluctuations of the sum $S_n$ that are of order $O(n)$ with respect to $S_n$ or, equivalently, order $O(1)$ with respect to the sample mean $S_n/n$. These fluctuations are the focus of the theory of large deviations \cite{touchette2009,harris2013,ellis1985,oono1989,ellis1995,ellis1999} and are known to be characterized by the following exponential pdf:
\begin{equation}
\Prob( S_n/n = s ) = e^{- n I(s) +o(n) }
\label{eq:ldp1}
\end{equation}
where $o(n)$ denotes corrections growing slower than linearly in $n$. We say that $S_n/n$ satisfies a \emph{large deviation principle} (LPD) if its pdf has the form above or, equivalently, if the following limit exists:
\begin{equation}
I(s) = \lim_{n\rightarrow\infty} -\frac{1}{n}\ln \Prob( S_n/n = s ).
\end{equation}
The function $I(s)$ defined by this limit is called the \emph{rate function}; it governs according to (\ref{eq:ldp1}) the rate at which $\Prob(S_n/n)$ decays to 0 when $n\rightarrow\infty$ and so the rate at which this pdf concentrates exponentially with $n$ around the typical values of $S_n/n$ corresponding to the zeros of $I(s)$. This will be studied in more detail below.

The main result that we will use to study the large deviations of $S_n/n$ is the G\"artner-Ellis Theorem \cite{gartner1977,ellis1985}, which enables one to obtain the rate function $I(s)$ from the so-called \emph{scaled cumulant generating function} (SCGF) defined as
\begin{equation} \label{eq:LDT:lambda}
\lambda(w) = \lim_{n\rightarrow\infty} \frac{1}{n}\ln g_n(w), 
\end{equation}
where 
\begin{equation} \label{eq:Def:gn}
g_n(w) = \Esp{e^{w S_n}}=\int_\R dx_1\cdots\int_\R dx_n\, \Prob(x_1,\ldots,x_n)\, e^{w\sum_{i=1}^n x_i}
\end{equation}
is the generating function of $S_n$. The G\"artner-Ellis Theorem states in simplified form that, if $\lambda(w)$ exists and is differentiable everywhere, then $S_n/n$ satisfies an LDP with rate function $I(s)$ given by the Legendre-Fenchel transform of $\lambda(w)$: 
\begin{equation}
I(s) = \lambda^\star(s) = \sup_{w} \{w s - \lambda(w)\}.
\end{equation}

In the next section, we will see that the differentiability property of $\lambda(w)$ is not always satisfied, a sign that $I(s)$ is either nonconvex or has linear parts \cite{touchette2009}. Interestingly, both cases can arise for sums of matrix-correlated random variables, as will be shown with explicit examples in Sec.~\ref{ELD}, and must be treated with a local version of the G\"artner-Ellis Theorem or other results \cite{touchette2009}, such as the complex integral method described in \cite{touchette2010}. The reason why this theorem does not apply for these cases has to do essentially with the fact that a nonconvex $I(s)$ and its convex hull have the same $\lambda^\star(s)$, which is nondifferentiable somewhere; see Sec.~4.4 of \cite{touchette2009} for more details.

\section{Large deviation results }
\label{sec:GEMR}

The G\"artner-Ellis Theorem takes a simple form for $\iid$ random variables due to the fact that the generating function $g_n(w)$ factorizes into a product of identical marginal generating functions, $g(w)=\Esp{e^{wX}}$, so that 
\begin{equation}
\lambda(w)=\lim_{n\rightarrow\infty}\frac{1}{n}\ln g(w)^n=\ln g(w).
\label{scgfiid1}
\end{equation}
Using the product structure of the matrix product ansatz, we show in this section that a similar result holds for matrix-correlated random variables by replacing $g(w)$ with a matrix generating function $\mG(w)$. Specific examples of rate functions obtained from this matrix generalization of the G\"artner-Ellis Theorem are presented in the next section.

\subsection{Scaled cumulant generating function}

Since $\Lf$ is a linear form, we can directly expand (\ref{eq:Def:gn}) in the definition of the generating function $g_n(w)$ to obtain
\begin{equation} \label{eq:GEMR:gn}
g_n(w) = \int_\R \frac{1}{\Lf\p{\mE^n }} \Lf\p{ e^{w x_1} \mR(x_1) dx_1 \dots e^{w x_n} \mR(x_n) dx_n)   }= \frac{ \Lf\p{ \mG^n(w)} }{ \Lf\p{ \mE^n } },
\end{equation}
where
\begin{equation}
\mG(w) \equiv \int_\R e^{w x}\mR(x) dx
\end{equation}
is the matrix generating function of the $\mR$ matrix. Each component of $\mG_{i,j}(w)$ can be decomposed as
\begin{equation} \label{eq:mG:decomposition}
\mG_{i,j}(w)= \mE_{i,j} g_{i,j}(w)
\end{equation}
where $g_{i,j}(w)$ is the generating function associated with the well defined (positive and normalizable) pdf $\mP_{i,j}(x)$ defined in (\ref{Pdef1}). Substituting (\ref{eq:GEMR:gn}) in the limit defining the SCGF, we then obtain
\begin{equation} \label{eq:GEMR:lambda:0}
\lambda(w)= \lim_{n\rightarrow\infty} \frac{1}{n}\ln \Lf\p{\mG(w)^n} - K(\mE),
\end{equation}
where
\begin{equation} \label{eq:K}
K(\mE) = \lim_{n\rightarrow\infty} \frac{1}{n}\ln \Lf\p{\mE^n},
\end{equation}
is a constant independent of $w$. In fact, this constant can be taken to be $0$, since for any structure matrix $\mE$ there is a constant $\eta > 0$ such that $K(\eta \mE)=0$ and the pdf~(\ref{eq:def}) is invariant by scalar multiplication of the structure matrix. The SCGF is given accordingly only by the limit involving the matrix $\mG(w)$.

Our result (\ref{eq:GEMR:lambda:0}) for the SCGF has an obvious similarity with the product structure of the $\iid$ result of (\ref{scgfiid1}), as well as with the case of sample means of Markov-correlated random variables, for which we have
\begin{equation}
\lambda(w)= \lim_{n\rightarrow\infty} \frac{1}{n}\ln \Tra \rho\pi(w)^{n-1},
\end{equation}
where $\pi(w)_{i,j}=\pi_{i,j}e^{w j}$ is the so-called tilted matrix obtained from the transition matrix $\pi_{i,j}$ of the (assumed ergodic) Markov chain, $\rho$ is the initial pdf of the Markov chain and $\Tra v= \sum_i v_i$ is the vector component sum \cite{touchette2009}. The similarity with Markov chains stems directly from the hidden Markov representation mentioned earlier. However, it is important to note that $\pi(w)$ and $\mG(w)$ arise from different contexts and follow different sets of constraints. First, $\pi(w)$ is not properly speaking a generating function whereas $\mG(w)$ is. Moreover, $\pi(0)=\pi$ is a stochastic matrix (i.e., $\sum_j \pi_{i,j}=1$), whereas $\mG(0) = \mE$ is only a non-negative matrix. The matrix function $\mG(w)$ is therefore less constrained than $\pi(w)$ in general. For instance, the dominant Jordan blocks $J_{1,k}$ of $\pi$ are necessarily of size $1$ because $\pi$ describes a probability flow, which has to vanish between different stationary states. The structure matrix $\mE$ is not subject to this restriction because it represents an affinity (i.e., a non-normalized probability) rather than a probability flow.

To find the SCGF $\lambda(w)$, we expand $\mG(w)^n$ using its Jordan decomposition. We slightly extend our previous notation to now denote by $J_k(w)$ the Jordan block associated with the $m(w)$ distinct eigenvalues $\nu_k(w)$ of $\mG(w)$. Then
\begin{equation}
\mG(w)= \sum_{k=1}^{m(w)} \tilde{J}_k(w) = \sum_{k=1}^{m(w)} (\tilde{J}_k(w) -  \nu_k(w) \mId) +  \nu_k(w) \mId
\end{equation}
where $\tilde{J}_k(w)$ is the Jordan matrix $J_k(w)$ completed with zeros outside of the block $k$: 
\begin{equation}
\tilde{J}_k(w) = B \begin{pmatrix} 0 \hspace{3pt} \raisebox{-10pt}{$\ddots$}  &  &0 \\ & { J_k(w) } & \\ 0 & & \raisebox{7pt}{$\ddots$} \hspace{3pt}0  \end{pmatrix}  B^{-1}.
\end{equation}
Using $[\mId, \tilde{J}_k]=0$ and 
\begin{equation}
(\tilde{J}_k(w) -  \nu_k(w) \mId)^d=0,
\end{equation}
 we can expand $\mG(w)^n$ as
\begin{equation} \label{eq:GEMR:mGn}
\mG(w)^n = \sum_{k=1}^{m(w)} \sum_{r=1}^{d} \nu_k(w)^n  \binom{n}{r} \p{\frac{\tilde{J}_k(w)}{\nu_k(w)} - \nu_k(w) \mId}^r.
\end{equation}
Substituting (\ref{eq:GEMR:mGn}) in (\ref{eq:GEMR:lambda:0}) then leads to
\begin{equation} \label{eq:GEMR:lim}
 \lim_{n\rightarrow\infty} \frac{\ln \Lf\p{ \mG(w)^n} } {n} = \max_l \{ \ln |\nu_l(w)| \},
\end{equation}
where the maximum is on the eigenvalues $\nu_l(w)$ such that $\Lf( \tilde{J}_l(w) )\ne0$. This condition eliminates eigenvalues coming from $\mE$ that are unreachable in the hidden chain $\Gamma$ due to a particular choice of $\mA$. The conclusion that we reach from (\ref{eq:GEMR:lim}) is that the limit defining $\lambda(w)$ exists for any matrix representation $\mR(x)$ and reads
\begin{equation} 
\label{eq:GEMR:lambda:max}
\lambda(w) = \max_l \{ \lambda_l(w) \}=\ln \nu_1(w),
\end{equation}
with $\lambda_l(w) = \ln |\nu_l(w)|$.
This result is similar to the Markov chain case, for which $\lambda(w)$ is given by the dominant eigenvalue of the matrix $\pi(w)$. Thus, although $\pi(w)$ and $\mG(w)$ are rather different matrices, they yield similar SCGFs in the $n\rightarrow\infty$ limit because of the Perron-Frobenius Theorem. In both cases, a lot of structure contained in the generating function is in fact lost in the SCGF.

\subsection{Rate function}

To apply the G\"artner-Ellis Theorem, we need to study the differentiability of $\lambda(w)$. As in the Markov case, if the dominant eigenvalue is unique, then $\lambda(w)$ is as smooth as $\mG(w)$. However, if we have an eigenvalue collision for some $w$, i.e., if for some $w$, say $w=w_0$, the dominant eigenvalue $\nu_1(w_0)$ has a multiplicity greater than 1, then $\lambda(w_0)$ may be nondifferentiable because of the maximum in (\ref{eq:GEMR:lambda:max}).
Physically, this non-differentiability is interpreted as a dynamical first-order phase transition at $w_0$ \cite{AppertRolland2008,Lefevere2010,Lefevere2011,Lecomte2012}. Around such a collision point, the colliding log-eigenvalues $\lambda_r(w)$ of $\mG(w)$ can be generically approximated as 
\begin{equation} \label{eq:EigC:st}
\lambda_r(w_0+w)  =  \lambda(w_0) + \mu_r w +O(w^2),
\end{equation}
so that, by (\ref{eq:GEMR:lambda:max}), we have
\begin{equation}
\lambda(w_0+w) = \lambda(w_0) + 
\begin{cases} 
\max \{ \mu_r \}\, w +O(w^2) 	& \text{if }w >0 \\ 
\min \{\mu_r\}\, w + O(w^2)	& \text{if } w <0 .
\end{cases} 
\end{equation}
Thus, we see that if the $\mu_r$ are different, then $\lambda(w)$ is nondifferentiable at $w_0$. In this case, we can still apply the G\"artner-Ellis Theorem but only locally at points where $\lambda(w)$ is differentiable; see Sec. 4.4. of \cite{touchette2009} for more details. Here, this means that we can apply this theorem at all $w$ except $w_0$. Doing so yields $I(s)$ as the Legendre transform of $\lambda(w)$ for $s\leq\min\mu_r$ and $s\geq\max\mu_r$, but not for $s\in (\min \mu_r,\max \mu_r)$ because of the nondifferentiability of $\lambda(w)$ at $w_0$ and the fact that $\lambda'(w_0-0)=\min \mu_r$ and $\lambda'(w_0+0)=\max \mu_r$. On this open interval, $I(s)$ can be nonconvex or linear, but this cannot be determined from $\lambda(w)$, as explained in more detail in \cite{ellis1995,touchette2009}.

Nonconvex rate functions typically appear in non-ergodic Markov chains \cite{dinwoodie1992,dinwoodie1993,ellis1999} in addition to mixtures of $\iid$ sample means \cite{ioffe1993,touchette2009}, whereas rate functions with linear branches are known to arise in Markov chains with absorbing states. All these cases can lead to extensions of the Law of Large Numbers involving more than one concentration points. In our case, we expect this sort of extensions to arise whenever long-range correlations are present, since this case of correlations appears when the structure matrix $\mE$ has a multiple dominant eigenvalue at $w=0$. This will be investigated in Sec.~\ref{ELD}.

\subsection{Connection with the Law of Large Numbers and the Central Limit Theorem} 
\label{sec:Conn}

The rate function $I(s)$ provides information not only about the large deviations of the sample mean $S_n/n$, but also about its small deviations and its most probable values corresponding to the zeros and global minima of $I(s)$. In the case where $I(s)$ has only one global minimum at $s=\mu$, then the most probable value is also the typical value, in the sense that $S_n/n$ converges almost surely to $\mu$. In this case, the sample mean thus concentrates to the mean, in accordance with the Law of Large Numbers.

Using the G\"artner-Ellis Theorem, we can express the concentration point $\mu$ as $\mu=\lambda'(0)$, assuming that $\lambda(w)$ is differentiable at $w=0$. In our case, we have
$\lambda(w) = \ln \nu_1(w)$ and the normalization condition $K(\mE)=0$ implies (see~(\ref{eq:K})) that $\nu_1(0)=1$ , so that $\lambda'(0)= \nu'_1(0)$.
Using classical perturbation theory \cite{kato1995} for the eigenvalue $\nu_1(w)$ then yields
\begin{equation}
\mu = \nu'_1(0) = \sum_{i,j} \rho_i \mG'_{i,j}(0) v_j
\end{equation}
where $\rho$ and $v$ are respectively the left- and right-eigenvectors of $\mE$ associated with $\nu_1(0)$, normalized so that $\tr \nu=1$ and $\nu^T \rho=1$. Since the matrix $\mG'(0)$ can be decomposed as
\begin{equation}
\begin{aligned}
\mG'_{i,j}(0) = \mE_{i,j} \p{ \partial_w \int_\R e^{-w x} \mP_{i,j}(x) dx}\Bigg{|}_{w=0}  = \mE_{i,j} \mu_{i,j} 
\end{aligned}
\end{equation}
where $\mu_{i,j}$ is the mean of the pdf $\mP_{i,j}(x)$, we also have
\begin{equation} \label{eq:GEMR:mu}
\mu = \sum_{i,j} \p{\rho_i \mE_{i,j} v_j} \mu_{i,j}.
\end{equation}
Therefore, we see that the concentration point of $S_n/n$ is a weighted average of the mean of $\mP_{i,j}(x)$. This should hold in general whenever $\mE$ leads to short-range correlations.

As before, we can understand this result probabilistically by appealing to the hidden Markov chain representation. For collision-free structure matrices $\mE$, the hidden Markov chain is known to converge towards a stationary state almost surely \cite{Angeletti2013:MatSum:Letter,Angeletti2013:MatSum:Long}. In this state, the mean of $S_n/n$ is then a weighted mixture of the mean of different distributions $\mP_{i,j}(x)$, where the weights are the stationary probabilities, corresponding to $\rho_i \mE_{i,j} v_j$, of observing a transition from state $i$ to $j$ in the hidden Markov chain.

To close this section, let us study the fluctuations of $S_n$ around its concentration point, characterized by the behavior of $I(s)$ around its minimum. It is often stated that if $I(s)$ has a quadratic minimum, then the sum $S_n$ satisfies the Central Limit Theorem in the sense that 
\begin{equation}
\frac{S_n - \lambda'(0) n} { \sqrt{ \lambda''(0) n} }
\end{equation}
converges in distribution towards the normal distribution. However, this relation is not rigorously valid: to obtain the Central Limit Theorem requires further conditions, such as $\lambda(w)$ to be holomorphic \cite{bryc1993}.

In our case, $\lambda(w)$ is holomorphic if $\mG(w)$ is collision-free and holomorphic. The matrix function $\mG(w)$ is holomorphic whenever the cumulant generating functions of the pdf $\mP_{i,j}(x)$ are holomorphic. This holomorphism condition is stronger than the differentiability condition needed to derive the rate function $I(s)$  which is nevertheless satisfied for a large class of $\mP_{i,j}(x)$ \footnote{  Log-normal distributions are a noteworthy example of distributions which do not satisfy this condition, even if sums of log-normal $\iid$ random variables do converge towards a normal distribution.}. In the presence of a collision, it is possible to obtain a SCGF that is differentiable but non-holomorphic in $0$.
To see this, assume that the colliding log-eigenvalues $\lambda_r(w)$ are twice differentiable and $\lambda_r'(0)=0$. Then
\begin{equation}
\lambda_r(w) = \frac{\sigma_r^2} {2} w^2 + O(w^3)
\end{equation}
and
\begin{equation}
\label{eq:GEMR:lambda:mult:2} 
\lambda(w) = \frac {\max \{ \sigma_r^2 \}} {2} w^2 + O(w^3),
\end{equation}
so that $\lambda(w)$ is differentiable at $0$. However,  
considering now the argument of the function $\lambda_r(w)$ as a complex  
variable z and rewriting these expressions in the direction $z=\imath \R$ yields
\begin{equation}
\lambda_r( z=\imath w) = -\frac {\sigma_r^2} {2} w^2 + O(w^3)
\end{equation}
and
\begin{equation}
\label{eq:GEMR:lambda:mult:2:bis} 
\lambda(z = \imath w) = - \frac{ \min\{ \sigma_r^2 \}}{2} z^2 + O(z^3),
\end{equation}
showing that $\lambda(w)$ is not holomorphic if the constants $\sigma_r$ are not equal. In this case, the Central Limit Theorem might not hold for the rescaled sum $S_n/\sqrt{n}$, consistently with the results of \cite{Angeletti2013:MatSum:Letter}.

\section{Long-range correlation examples}
\label{ELD}

We now give illustrations of the two most interesting cases of correlation obtained with the matrix ansatz, namely, polynomial and constant correlations, and obtain the rate functions of the sample mean $S_n/n$ for both cases. For simplicity, we  
consider the case of two-dimensional matrices (d=2), which already shows interesting large deviation behavior.

\subsection{Rate function with flat part} 
\label{sec:ILD:flat}

For $d=2$, the only structure matrix (up to some trivial transformations) leading to polynomial correlation is $\mE=\mId+U$ \cite{Angeletti2013:MatSum:Long}. This kind of correlation structure does not exist for finite Markov chain, so it is interesting to determine the rate function $I(s)$ in this case.

For simplicity, we consider the projection matrix
\begin{equation}
\mA= 
\begin{pmatrix}
 0 & 1 \\ 
 0 & 0 
\end{pmatrix},
\end{equation}
for which the generating function $g_n(w)$ can be computed exactly as
\begin{equation} \label{eq:ELD:Il:gn}
g_n(w) = \mG_{1,2}(w) \frac{ \mG_{1,1}^n(w) - \mG_{2,2}^n(w)}{\mG_{1,1}(w) - \mG_{2,2}(w)},
\end{equation}
where $\mG_{i,j}(w)$ represents the (scalar) generating function of the pdf $\mP_{i,j}(x)$.
Due to the choice of $\mE$, the matrix $\mG(w)$ is triangular superior. 
The diagonal coefficients $\mG_{k,k}$ are therefore the eigenvalues $\nu_k(w)$ of $\mG(w)$. 
Defining $c(w)= \mG_{1,2}(w)$ and using the notation $\lambda_k(w) =  \ln \nu_k(w)$ introduced in the previous section then yields
\begin{equation} \label{eq:ELD:isgn}
 g_n(w)	= c(w) \frac{e^{ n \lambda_1(w) } - e^{ n \lambda_2(w) } }{ e^{  \lambda_1(w) } - e^{ \lambda_2(w) } },
\end{equation}

The expression of $g_n(w)$ is at this point quite general and covers many cases of large deviations. To be more concrete, we now make a number of assumptions leading to a rate function having a flat part. For this, we assume that
\begin{equation}
\lambda'_1(0)=\mu_1 < \lambda'_2(0) = \mu_2
\end{equation} 
and that $\lambda_1(w) = \lambda_2(w)$ has the only solution $w=0$. In this case,  it can be verified from (\ref{eq:ELD:isgn}) that $\lambda(w)$ is differentiable everywhere except at $w=0$. The left- and right-derivatives of $\lambda(0)$ are $\lambda'(0^-)=\mu_1$ and $\lambda'(0^+)=\mu_2$, respectively. Following our discussion of the G\"artner-Ellis Theorem of the previous section, the rate function $I(s)$ can therefore be obtained from the Legendre transform of $\lambda(w)$ for $s\in (-\infty,\mu_1] \cup [\mu_2,\infty)$ but not for the complementary interval $(\mu_1,\mu_2)$. In the latter interval, $I(s)$ could be flat or nonconvex, but as mentioned before the knowledge of $\lambda(w)$ is not sufficient to discriminate between these two cases.

However, since we have the exact expression of the generating function for any $n$ and therefore more information than the sole limit $\lambda(w)$,
we can obtain the pdf of $S_n/n$ by inverting the Laplace transform
\begin{equation}
P(S_n/n=s) = \frac 1 {2 \imath \pi }\int_{r - \imath \infty}^{r+ \imath \infty} g_n(w) e^{ - n w s} dw,
\end{equation}
where $r$ is an arbitrary real constant in the region of convergence of $g_n(w)$. From this integral, involving the so-called Bromwich contour, we can obtain the rate function of $S_n/n$ following the method proposed in \cite{touchette2010} by expanding $g_n(w)$ in series form and by applying a saddlepoint approximation to each term in the series, taking care of any pole when transforming the integration contour to reach the saddlepoints.

In our case, the series is simply $g_n(w)= g_{n,1}(w) - g_{n,2}(w)$, where
\begin{equation}
g_{n,k}(w) = c(w) \frac{ e^{ n \lambda_k(w) } } { e^{  \lambda_1(w) } - e^{ \lambda_2(w) } }.
\end{equation}
The two functions $g_{n,1}(w)$ and $g_{n,2}(w)$ have a pole in $0$, since $\lambda_1(0) = \lambda_2(0) = 0$, whose residue is
\begin{equation}
\Res_{g_{n,1}(w) } (0) = \frac{1}{\mu_1 - \mu_2} = \Res_{g_{n,2}(w)} (0).
\end{equation}
With this, we now split the integral of the inverse Laplace transform in two parts:
\begin{equation}
P(S_n/n=s) = p_{n,1}(s) - p_{n,2}(s),
\end{equation}
where
\begin{equation}
p_{n,k}(s) = \frac 1 {2 \imath \pi } \int_{r - \imath \infty}^{r+ \imath \infty} g_{n,k}(w) e^{ - n ws } dw,
\end{equation}
and apply the saddlepoint approximation to each term, starting with $r\neq 0$, by deforming the Bromwich contour to go through the saddlepoint $\Col{k}$ of the exponential term satisfying
\begin{equation} 
\lambda_k'(\Col{k}) = s.
\label{eq:saddle}
\end{equation}
Two different situations then arise:
\begin{itemize}
\item If $\Col{k}$ has the same sign as $r$, the integration path can be deformed
without going through the pole at $w=0$, which leads to
\begin{equation}
p_{n,k}(s)  \approx e^{ - n [ s \Col{k} - \lambda_k(\Col{k}) ] } = e^{- n \lambda_k^{\star}(s)}
\end{equation}
where $\lambda_k^\star$ is the Legendre transform of $\lambda_k$. 
\item The deformation of the contour to the saddlepoint must cross the pole $0$, in which case the residue must be included:
\begin{equation}
p_{n,k}(s) \approx \Res_{g_{n,k}(w)}(0) + e^{- n \lambda_k^{\star}(s)}.
\end{equation}
\end{itemize}

Here, the transition between these two regimes happens for $s=\mu_k$ since $\lambda'_k(0) = \mu_k$. For instance, if we choose $r<0$, then
\begin{equation}
p_{n,k}(s) \approx e^{- n \lambda_k^{\star}(s)} + 
\begin{cases}
 0 &  s < \mu_k,\\
 \Res_{g_{n,k}(w)}(0) & \mu_k < s.  \\
\end{cases}
\end{equation}
Outside the interval $[\mu_1,\mu_2]$, the residue terms cancel each other, so that combining $p_{n,1}(s)$ and $p_{n,2}(s)$ yields
\begin{equation}
I(s) =
\begin{cases}
 I_1(s) & s < \mu_1,\\
 0 	& \mu_1 < s < \mu_2,\\
 I_2(s) & \mu_2 < s.
\end{cases}
\end{equation}
Here $I_k(s)= \lambda_k^{\star}(s)$ actually corresponds to the rate function of a sample mean $S_n/n$ of $\iid$ random variables distributed with pdf $\mP_{k,k}(x)$.

\begin{figure}[t]
\centerline{ \includegraphics[width=8cm]{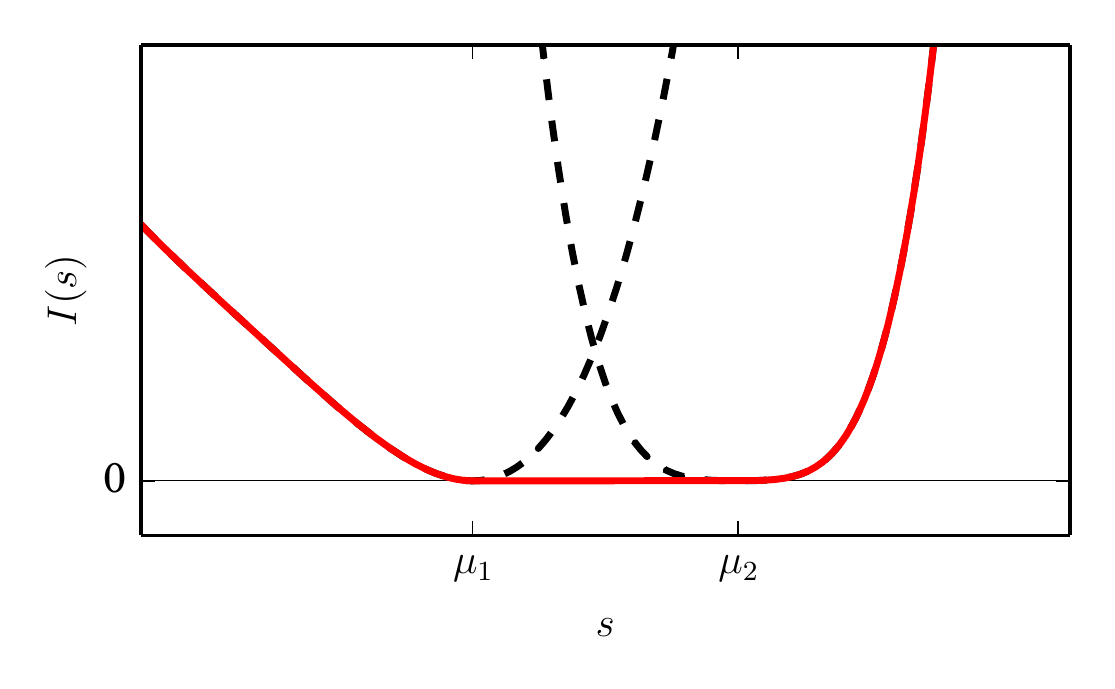} }
\caption{\label{fig:rate:flat} Sketch of the rate function for linear irreversible structure matrix.
Dashed lines: underlying rate function $I_1(s)$ and $I_2(s)$. Red solid line: resulting rate function $I(s)$.}
\end{figure}

The result for $I(s)$ is sketched in Fig.~\ref{fig:rate:flat}. As clearly seen, the effect of the pole in each term of the generating function is that $I(s)$ is flat for $x\in (\mu_1, \mu_2)$. This automatically implies that there is no Law of Large Numbers for $S_n/n$, that is, $S_n/n$ does not concentrate to a Dirac-delta pdf in the limit $n\rightarrow\infty$. It may instead converge to a pdf that scales slower than exponentially for $s\in(\mu_1,\mu_2)$ or to some stationary distribution that does not scale with $n$.

To find out, we can use the hidden Markov representation. For the structure matrix $\mE= \mId + U$, the corresponding hidden Markov chain is a highly non-stationary Markov chain \cite{Angeletti2013:MatSum:Letter}, which for $d=2$ stays in its initial state for a random time uniformly distributed on $\{0, 1,\dots, n\}$ before jumping to its final state. Consequently, the Markov chain does not converge almost surely towards a stationary state, and in this case, it can be shown that the sample mean actually converges towards the uniform distribution on the interval $[\mu_1, \mu_2]$ \cite{Angeletti2013:MatSum:Letter}. Thus, we have a concentration phenomenon on a whole interval rather than on a point, explaining the flat branch in the rate function $I(s)$.

Our particular choice for the matrices $\mE$ and $\mA$ simplifies the computations leading to this flat branch, but is otherwise not significant. Using the Perron-Frobenius decomposition of the matrix $\mE$ into irreducible blocks, it is possible to show that $g_n(w)$ has a form similar to (\ref{eq:ELD:Il:gn}) whenever $\mE$ leads to long-range correlations, so that the calculation steps given above are representative of this case. Consequently, we expect flat branches in rate functions to be a generic phenomenon for matrix-correlated random variables $X_1,\ldots, X_n$ exhibiting polynomial long-range correlation.

\subsection{Nonconvex rate function}

We now consider a model with constant correlation. For $d=2$, the only structure matrix with such correlation is
\begin{equation}
\mE = \begin{pmatrix} 1 & 0\\ 0 & 1 \end{pmatrix}.
\end{equation}
Choosing, for simplicity,
\begin{equation}
\mA = \begin{pmatrix} \frac 1 2 & 0\\ 0 & \frac 1 2 \end{pmatrix},
\end{equation}
the computation of the generating function is then straightforward and leads to
\begin{equation} \label{eq:ELD:NCg}
g_n(w) = \frac{1}{2} \p{ e^{  n \lambda_1(w) } + e^{  n \lambda_2(w) } }.
\end{equation}
As before, we have $\lambda_1(0)=\lambda_2(0)=0$ and the SCGF is not differentiable assuming $\lambda'_1(0) \ne \lambda'_2(0)$. In this case, we calculate the rate function using the inverse Laplace transform method, and, since there is now no pole, the calculation is much simpler and leads to 
\begin{equation}
I(s) = \min \{ I_1(s), I_2(s) \}
\end{equation}
where $I_k(s)= \lambda_k^{\star}(s)$ corresponds again to the rate function of a sample mean of $\iid$ random variables with pdf $\mP_{k,k}(x)$.

\begin{figure}[t]
\centerline{ \includegraphics[width=8cm]{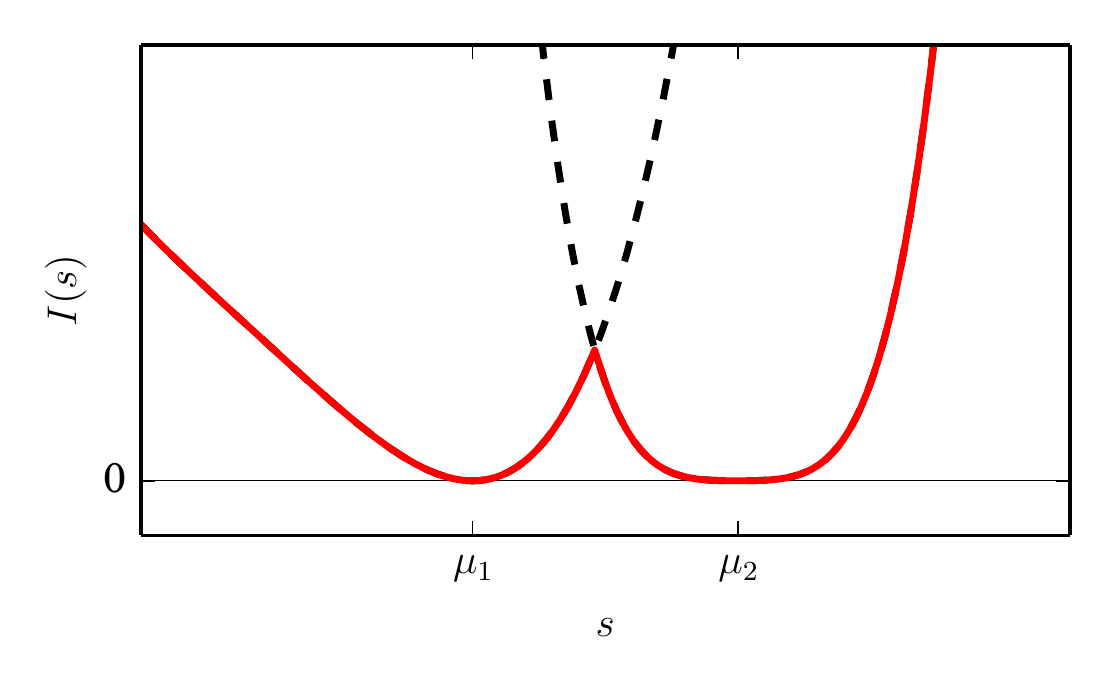} }
\caption{\label{fig:rate:nc} Sketch of the rate function for identity structure matrix.
Dashed lines: underlying rate function $I_1(s)$ and $I_2(s)$. Red solid line: resulting rate function $I(s)$.}
\end{figure}

The full rate function $I(s)$ is sketched in Fig.~\ref{fig:rate:nc}. Its main feature is that it is nonconvex with our assumption that $\lambda'_1(0) \ne \lambda'_2(0) $, which implies for this case that there are two distinct concentration points for $S_n/n$. These two concentration points can be understood again using the hidden Markov chain representation. For the structure matrix $\mE=\mId$, the hidden Markov chain stays in the same state from its beginning to its end. The concentration and fluctuations of $S_n/n$ therefore simply depend on the choice of initial state, similarly to the case of non-ergodic Markov chains.

This example is simple, but should be relevant to other cases of matrix-correlated variables leading to constant correlation. In this case, it is possible to show, using the Perron-Frobenius decomposition of the matrix $\mE$ mentioned before, that decoupled terms such as those appearing in the right-hand side of (\ref{eq:ELD:NCg}) are present in the expression of $g_n(w)$. As these decoupled terms are responsible for the nonconvex part of $I(s)$, we expect nonconvex rate functions to be generic in this case.

\section{Conclusion}

We have shown in this contribution how to obtain large deviations for sums of random variables described by a matrix product ansatz. These random variables are also described by a complementary hidden Markov model, which is probably the more natural setting for studying limits laws, such as the Law of Large Numbers and Central Limit Theorem, as was done in \cite{Angeletti2013:MatSum:Letter}. However, as shown here, the matrix product representation offers a natural starting point for generalizing large deviations results of $\iid$ variables in the context of the G\"artner-Ellis Theorem by introducing a matrix cumulant generating function. 

For system with short-range correlation, the G\"artner-Ellis Theorem was used to show that analogues of the Law of Large Number or the Central Limit Theorem hold. In the presence of long-range correlation, the direct calculation of the rate function for two specific examples has shown that flat and nonconvex rate functions are possible. Nonconvex rate functions are associated with constant correlation and are similar in nature to nonconvex rate functions appearing in non-ergodic Markov chains. By contrast, polynomial correlation leads to flat rate functions, which cannot appear for sample means defined on finite Markov chains. 

\section*{Acknowledgments}

HT thanks Stefano Ruffo and Thierry Dauxois for supporting a visit to ENS Lyon with the ANR grant LORIS (ANR-10-CEXC-010-01).

\section*{References}
\bibliographystyle{unsrt}
\bibliography{MathProba,angeletti,MathLinAlg,hugo,ASEP,Pertubation,referee}

\end{document}